\documentclass{emulateapj}

\newcommand{\kms}          	{\ensuremath{{\rm km~s^{-1}}}} 
 
\newcommand{\cc}         		{\ensuremath{{\rm cm^{-3}}}} 
\newcommand{\cs}         		{\ensuremath{{\rm cm^{-2}}}} 
\newcommand{\jyb}          		{\ensuremath{{\rm Jy~beam^{-1}}}}

\newcommand{\tex}{\ensuremath{T_{\rm ex}}} 
\newcommand{\trot}{\ensuremath{T_{\rm rot}}}

\newcommand{\tkin}{\ensuremath{T_{\rm kin}}}

\newcommand{\ci}{\ion{C}{1}} 
\newcommand{\cii}{\ion{C}{2}}
\newcommand{\hi}{\ion{H}{1}} 
\newcommand{\NcNco}{\ensuremath{N_{\rm C}/N_{\rm CO}}} 
\newcommand{\Nc}{\ensuremath{N_{\rm C}}} 
\newcommand{\Nco}{\ensuremath{N_{\rm CO}}} 
\newcommand{\fc}{\ensuremath{f_{\rm c}}} 
\newcommand{\pksfull}{PKS 1830$-$211} 
\newcommand{\pks}{PKS 1830} 

\shorttitle{\ci\ in \pksfull\ with the eSMA} 
\shortauthors{Bottinelli et al.}

\begin{document} 

\title{Detection of \ci\ in absorption towards \pksfull\ with the eSMA}

\author{
{Sandrine~Bottinelli\altaffilmark{1}},
{A.~Meredith~Hughes\altaffilmark{2}},
{Ewine~F.~van~Dishoeck\altaffilmark{1,3}},
{Ken~H.~Young\altaffilmark{2}},
{Richard~Chamberlin\altaffilmark{4}},
{Remo~P.J.~Tilanus\altaffilmark{5,6}},
{Mark~A.~Gurwell\altaffilmark{2}},
{David~J.~Wilner\altaffilmark{2}},
{Huib~Jan~van~Langevelde\altaffilmark{7,1}},
{Michiel~R.~Hogerheijde\altaffilmark{1}},
{Robert~D.~Christensen\altaffilmark{8}},
{Hiroko~Shinnaga\altaffilmark{4}},
{Hiroshige~Yoshida\altaffilmark{4}} 
}

\altaffiltext{1}{Leiden Observatory, Leiden University, 
P.O. Box 9513, NL 2300 RA Leiden, The Netherlands;
{\tt sandrine@strw.leidenuniv.nl}.}
\altaffiltext{2}{Harvard-Smithsonian Center for Astrophysics, 
MS 42, 60 Garden Street, Cambridge, MA 02138
.} 
\altaffiltext{3}{Max-Planck-Institut f\"{u}r Extraterrestrische Physik, 
Postfach 1312, 85741 Garching, Germany.}
\altaffiltext{4}{Caltech Submillimeter Observatory Office, 
111 Nowelo St., Hilo HI 96720
.} 
\altaffiltext{5}{Joint Astronomy Center, 660 N. A'ohoku Place, 
University Park, Hilo HI 96720
.}
\altaffiltext{6}{Netherlands Organisation for Scientific Research, 
Laan van Nieuw Oost-Indie 300, NL-2509 AC The Hague, The Netherlands.}
\altaffiltext{7}{Joint Institute for VLBI in Europe, Radiosterrenwacht 
Dwingeloo, Postbus 2, NL-7990 AA Dwingeloo, The Netherlands
.} 
\altaffiltext{8}{Harvard-Smithsonian Center for Astrophysics, 
Submillimeter Array, 645 North A'ohoku Place, Hilo HI 96721
.}

\begin{abstract} 
We report the first science observations and results obtained with the ``extended''
SMA (eSMA), which is composed of the SMA (Submillimeter Array), JCMT 
(James Clerk Maxwell Telescope) and CSO (Caltech Submillimeter Observatory).
Redshifted absorptions at $z$=0.886 of
\ci\ ($^3P_1 - ^3P_0$) 
were observed with the eSMA
with an angular resolution of 0\farcs55$\times$0\farcs22 at 1.1~mm
toward the southwestern image of 
the remarkable lensed quasar \pksfull, but not toward
the northeastern component at a separation of $\sim1''$.
Additionally, SMA observations of CO, $^{13}$CO and C$^{18}$O (all $J$=4--3)
were obtained toward this object: CO was also detected 
toward the SW component, but none of the isotopologues were.
This is the first time [\ci] is detected in this object,
allowing the first direct determination of relative abundances 
of neutral atomic carbon to CO in the
molecular clouds of a 
spiral galaxy at $z>0.1$.
The [\ci] and CO profiles can be decomposed into two and three
velocity components respectively. 
We derive C/CO column density ratios ranging
from $\lesssim$0.5 (representative of dense cores) 
to $\sim$2.5 (close to translucent clouds values). 
This could indicate that we are seeing environments with
different physical conditions or that we are witnessing 
chemical evolution of regions where C has not completely been converted
into CO.
\end{abstract}

\keywords{quasars: individual (\pks) --- quasars: absorption lines 
--- galaxies: ISM --- ISM: abundances}

\section{Introduction} 
\label{sec:intro}

A powerful way to study the physical and chemical conditions in the 
interstellar medium (ISM) of 
distant galaxies is to observe chemical species in absorption against a strong 
continuum background source. 
Few such lines of sight have been unveiled so far but 
\objectname[QSO B1830-210]{\pksfull} (hereafter \pks) 
is one of the most remarkable systems allowing this kind of study. 
\pks\ is a radio-loud quasar with a redshift of $z$ = 2.507 
\citep{lidman-etal99}, for which
millimeter continuum emission images show two compact components, 
north-east (NE) and south-west (SW), separated by $\sim1''$
\citep{frye-etal97}. 
This double 
structure represents two images of the background quasar, 
magnified and distorted by a lensing system at $z$ = 0.88582  
\citep{wiklind+combes96}.  
This lens was at first indirectly detected by the observations of broad \hi\ and 
molecular absorptions at millimeter wavelengths
\citep{wiklind+combes96,wiklind+combes98,gerin-etal97,muller-etal06}. 
\citet{wiklind+combes98} suggested the lensing system to be an almost face-on
spiral galaxy, 
which was recently confirmed by direct optical and infrared HST images  
\citep{courbin-etal02,winn-etal02}. 

A key species in the physics and chemistry of the ISM in galaxies 
is atomic carbon.
For instance, [\ci] 492\,GHz is one of the major far-infrared (FIR) 
fine structure atomic lines via which
cooling of the gas occurs 
 \citep{hollenbach+tielens97}. 
 Moreover, both C and C$^+$, which directly follow from C,
are essential in radical reaction networks, 
especially those leading to carbon chains
\citep{suzuki-etal92,sternberg+dalgarno95}. 
Atomic carbon is also crucial for better understanding the structure 
of photo-dissociation 
regions (PDRs). 
Indeed, early observations of [\ci]  \citep[e.g.,][]{keene-etal85} 
revealed more intense emission than originally 
predicted by the models.  
Observations 
and theory now seem to have been reconciled by considering the effects of 
the clumpy structure of clouds 
 \citep[][and references therein]{hollenbach+tielens97}. 
\\
Several studies have investigated the use 
of the neutral atomic carbon (C) to CO 
abundance ratio (\NcNco) to
investigate the physical conditions of the gas. 
A number of such analyses make use of
UV absorption data, and hence are limited to diffuse clouds in our Galaxy.
In order to measure \Nc\ and \Nco\ in a dense environment,
many studies have targeted submillimeter emission of these species.
However, the lowest C transition lies at a rather high frequency
($^3P_1 - ^3P_0$ at $\sim$492~GHz), and so its detection requires 
the use of ground-based facilities with sensitive receivers and/or 
a large collecting area, 
or the observation of a redshifted source.
Moreover, in these submillimeter studies, 
the emission line data are averaged over large areas and suffer
from radiative transfer and excitation effects
(with a bias toward the densest regions). 
The eSMA
(``extended'' Submillimeter Array, \citealt{bottinelli-etal08-spie}) 
 is particularly well suited for this kind of observation as it provides
 the high sensitivity and high spatial resolution required to observe
 the weak [\ci] transition in \pks.
In this work, we present the first data obtained with the eSMA,
which probe
\NcNco\ directly for the first time in absorption in
the dense molecular regions of an external galaxy.

\section{Observations and data reduction} 

\subsection{eSMA and SMA data} 

The eSMA 
consists of the SMA\footnote{The Submillimeter Array is a joint project 
between the Smithsonian Astrophysical Observatory 
and the Academia Sinica Institute of Astronomy and
Astrophysics and is funded by the Smithsonian Institution and the  
Academia Sinica.} 
array of eight 6-m antennas augmented by the nearby
single dishes of the JCMT\footnote{The James Clerk Maxwell Telescope 
is operated by The Joint Astronomy Centre on behalf of the Science and 
Technology Facilities Council of the United Kingdom, 
the Netherlands Organisation for Scientific Research, 
and the National Research Council of Canada. } (15\,m) and the 
CSO (10.4\,m).  
The collecting area of the eSMA 
is twice that of 
the SMA alone, 
providing increased sensitivity, in particular  
for the longest baselines, which include the JCMT and CSO. 
Moreover, the longest baseline of the eSMA is increased by 
50\% compared to the SMA alone,
which can bring the angular
resolution down to below $0\farcs 3$ at 230~GHz, as reported here,
and to below 0\farcs 2 at 345\,GHz, which will be the preferred
frequency for eSMA observations.\\ 

Observations of \pks\ 
($\alpha_{\rm J2000}=18^{\rm h}33^{\rm m}39\fs 889$,  
$\delta_{\rm J2000}=-21^{\circ}03'39\farcs 77$) 
were carried out with the eSMA 
for seven hours on 2008 April 14 (with eight SMA antennas in the
very-extended configuration, or ``vex''), 
and for eleven hours on 2008 August 5 with 
the SMA alone, also in vex configuration. 
The eSMA observations targeted the \ci\ ($^3$P$_1-^3$P$_0$) transition
while the SMA provided CO, $^{13}$CO and C$^{18}$O  (all $J=4-3$) data. 
Table \ref{tab:obs} summarizes the main observational parameters.
The 45 eSMA baselines and the 28 SMA baselines had lengths ranging from 25-782\,m 
(22-678\,k$\lambda$)  and 68-509\,m (56-419\,k$\lambda$) respectively, 
resulting in a best resolution at 1.1~mm of 0\farcs55$\times$0\farcs22. 
The tracks interleaved 15 to 20-minute observations of \pks\ with 
3 to 5-minute observations of the calibrators.
Observations were carried out in single polarization mode
with half-wave plates in the beam of the CSO and JCMT that automatically
rotate to a common polarization as a function of elevation.

In this interim state of eSMA commissioning, the JCMT used receiver A3
which, due to a different intermediate frequency,
provided a usable bandwidth of
about 1.5~GHz, slightly less than the 2~GHz covered by the SMA and CSO.
For both eSMA and SMA tracks, the correlator was setup to provide
a uniform spectral resolution of 0.8125~MHz per channel across
the full 1.5-2~GHz bandwith, resulting in 2304 to 3072 channels per
baseline and velocity coverages
of $\sim$1725 and $\sim$2450~\kms, for the eSMA and SMA respectively. 
Weather conditions were poor on the two nights with
atmospheric opacity at 225~GHz ranging from 0.15 to 0.25.

\subsection{Data reduction} 

The data were reduced and calibrated using
the MIR/IDL
software package\footnote{\tt http://cfa-www.harvard.edu/$\sim$cqi/mircook.html}, 
and imaging was carried out with the MIRIAD software package.
The SMA receiver temperatures ranged 
from $\sim$45 to 70~K on both nights, with the CSO comparable
to the high end of this range, and the JCMT a factor of 2.5 times higher.
The IF passband phase and amplitude were calibrated with observations of the 
strong sources 3C273, 3C454.3, J1924$-$292, and NRAO530 for the eSMA observations,
and 3C273, 3C279 and 3C454.3 for the SMA-only data. 
Self-calibration of the eSMA phase of \pks\ was done on the shortest possible timescales 
using an initial model consisting of a pair of equal point sources separated 
by $1\farcs05$ at a position angle of $42^\circ$. 
The amplitude gains for
\pks\ were determined from the self-calibrated 
amplitudes of J1924$-$292.  
For the SMA-only data, phase calibration was performed in a standard way 
using J1733$-$130 to derive time-dependent amplitude gains throughout the night.
The amplitude gains imply average efficiencies of about 0.75 for the SMA 
antennas, 0.70 for the JCMT, and 0.30 for the CSO; the low efficiency of the 
CSO is likely due to a known malfunction of its rotating polarizer.
To set the flux scales, standard SMA monitoring 
observations of J1924$-$292 (eSMA) and 3C279 (SMA-only) 
at 230\,GHz and 345\,GHz were used  
(uncertainty $\lesssim 20$\%).


\section{Results} 
\label{sec:results}

\begin{figure}[ht]
\centering
\epsscale{.75}
\plotone{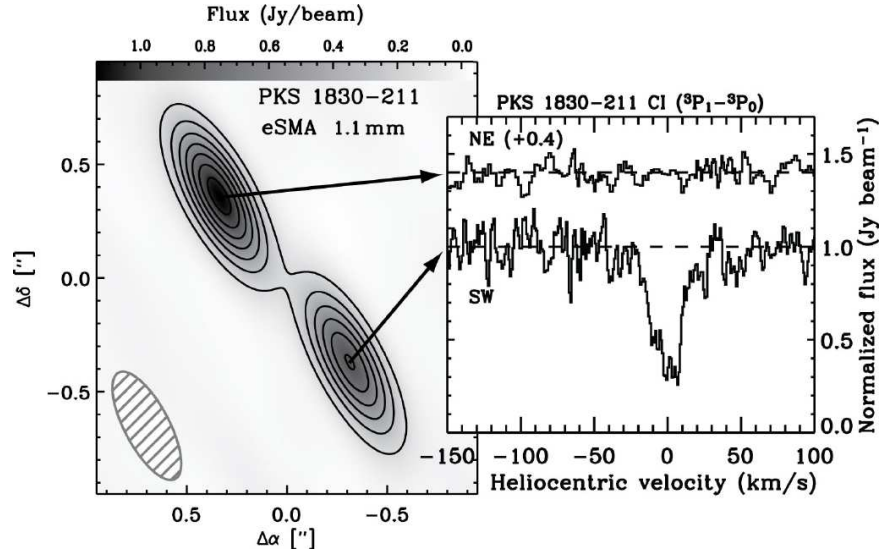}
\caption{eSMA 1.1~mm continuum map and [\ci] spectra of \pks. 
Absolute astrometry information is lost during the self-calibration process, 
so that the offsets here are relative to the center of the double point source.
The contour levels are multiples of 150~mJy. 
The 0\farcs55$\times$0\farcs22 beam is shown in the lower left corner.}
\label{fig:cont} 
\end{figure}

The continuum map, obtained with the eSMA and derived 
from line-free channels, is displayed in
Fig.~\ref{fig:cont}. This map shows 
the two lensed images of the distant quasar, separated by 1$''$,
consistent with previously reported values.
Figure~\ref{fig:cont} also shows the [\ci] spectra at the peak
of the continuum emission of 
the SW and NE components.

\begin{figure}[ht]
\centering
\includegraphics[bb=30 241 580 572,clip=true,width=0.55\columnwidth]{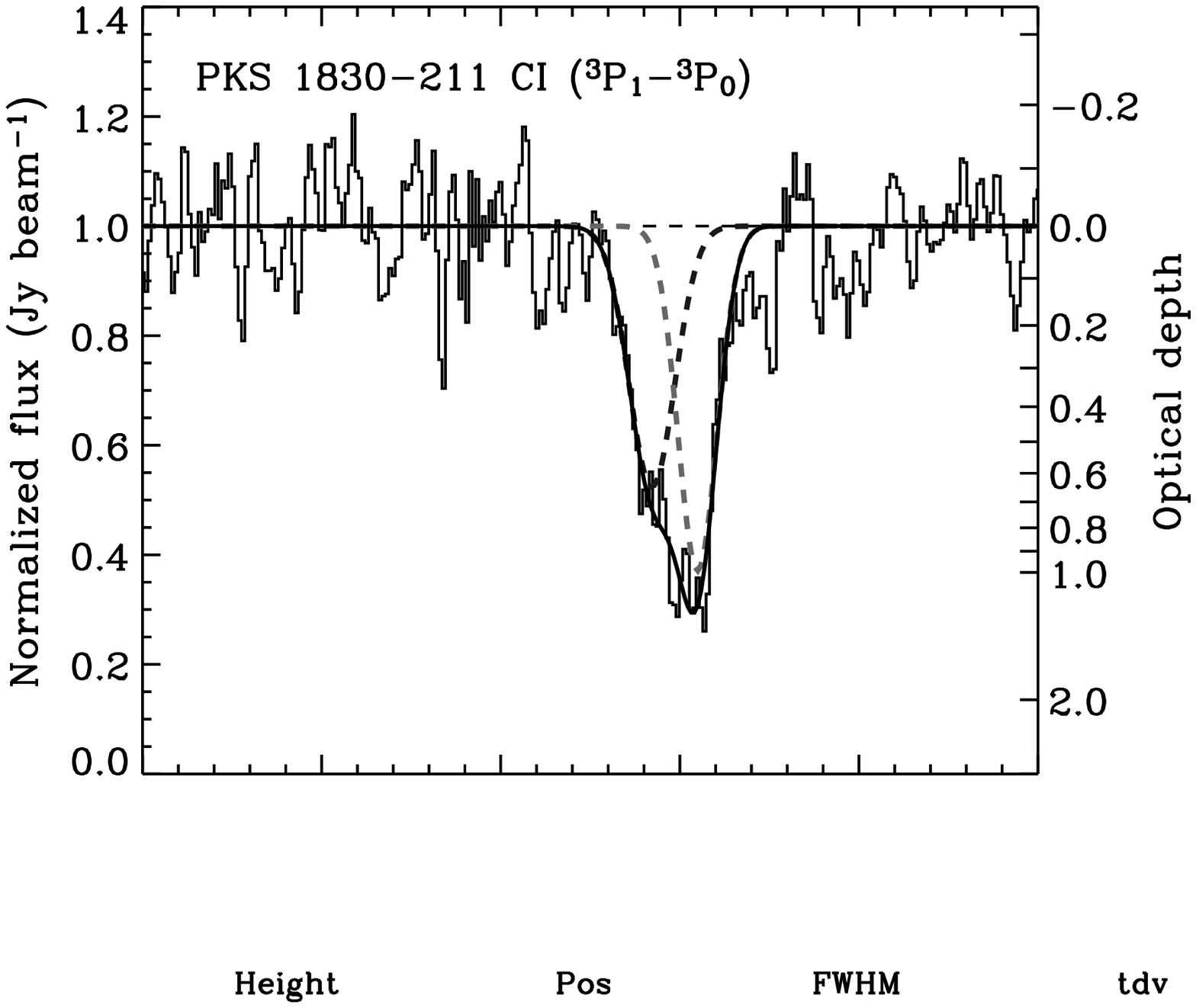} 
\includegraphics[bb=30 190 580 572,clip=true,width=0.55\columnwidth]{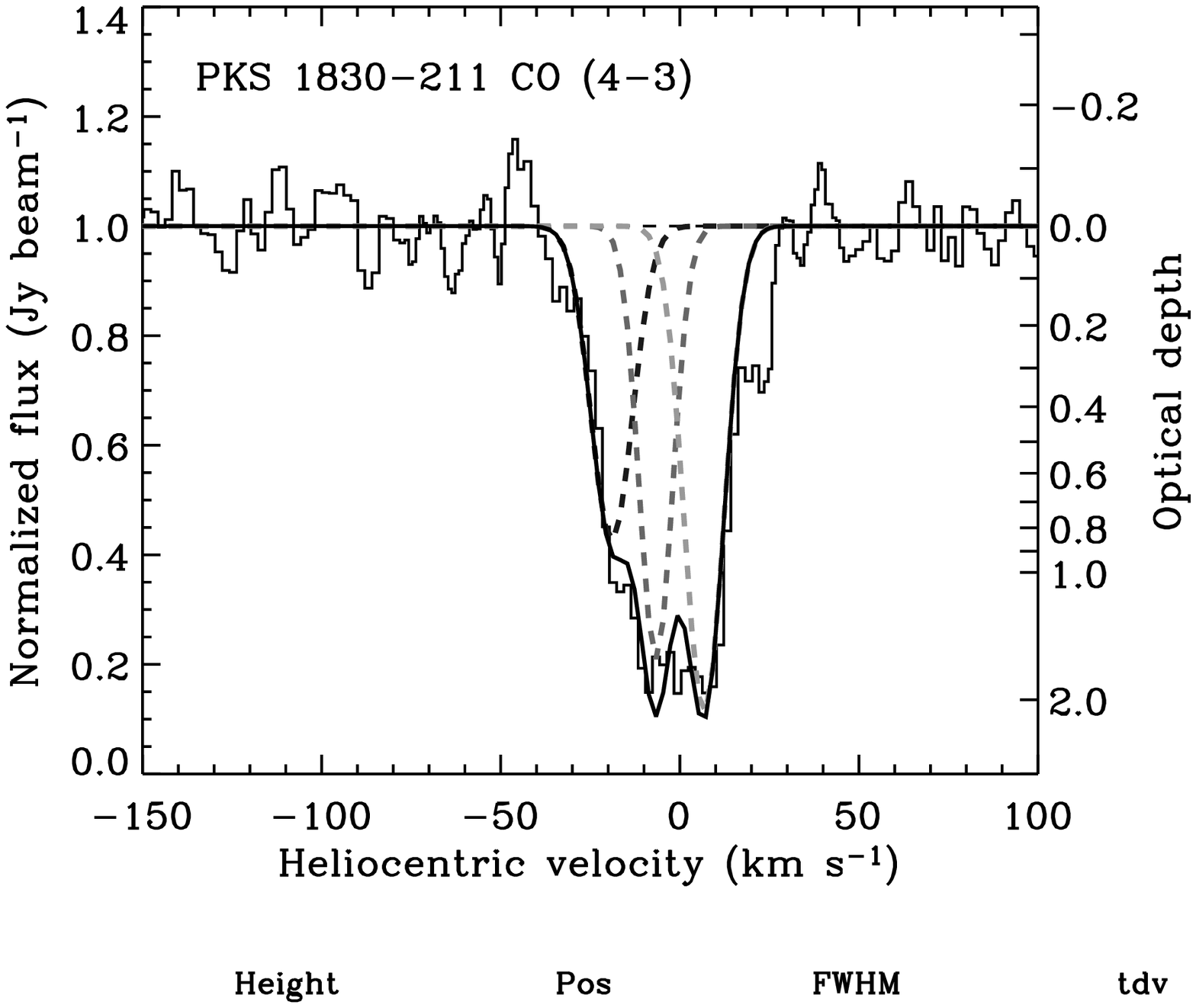}
\caption{eSMA spectrum of \ci\ ($^3P_1-^3P_0$) 
and SMA spectrum of CO ($J=4-3$) 
observed towards the SW component.
Spectra were smoothed to a velocity resolution of 2.8 and 5.9~\kms, yielding
rms before normalization of 0.07 and 0.04~\jyb,
corresponding to 0.09 and 0.05 on optical depth scale, for [\ci] and CO respectively.
Overlaid in dashed lines are two- and three-component 
gaussian fits to the 
[\ci] and CO profiles respectively, while
the solid lines are the sum of these individual components.
} 
\label{fig:spectra} 
\end{figure}

Figure~\ref{fig:spectra} shows the absorption spectra 
of [\ci]  and CO ($4-3$) as a function of heliocentric velocity, $V_{\rm hel}$,
observed in the SW source
by the eSMA and SMA respectively. 
Two- and three-component gaussian fits were applied to the
[\ci] and CO spectra respectively and are overplotted on Fig.~\ref{fig:spectra}.
Our [\ci] velocities and widths agree within the errors with those 
reported 
for other molecular 
species \citep[][]{muller-etal06,menten-etal08} and these
values were used to guide the fits to the blended CO components.
The presence of a fourth velocity component at $\sim$+20~\kms\
in the CO spectrum is possible, but was not fitted due to its
low signal-to-noise ratio and is therefore not investigated further in this work.
The derived parameters 
are given in columns 2-5 of Table~\ref{results tab}.
The optical depths at velocities $V_{\rm hel,0}$ were derived using
$\tau_0 = -\ln(F_{V_{\rm hel,0}}/F_{\rm c})$,
where $F_{V_{\rm hel,0}}$ and $F_{\rm c}$ are the line and continuum fluxes respectively.
This equation implicitly assumes that the covering factor \fc\ of the absorbing molecular
gas across the finite extent of the background continuum source is 1, 
which will be discussed in Section~\ref{sec:discussion}.
\\

Using line data from LAMDA, the Leiden Atomic and Molecular 
Database\footnote{\tt http://www.strw.leidenuniv.nl/$\sim$moldata/},
we ran the radiative transfer code RADEX \citep{vandertak-etal07}
for a range of H$_2$ number densities $n$ and kinetic temperatures \tkin, 
and for $T_{\rm CMB}$=5.16~K;
we obtain the excitation temperature for each ($n$, \tkin), and 
hence determine ($n$, \tkin)$\sim$($0.2-5~\times10^4$\,\cc, $10-30$\,K) corresponding to
the excitation temperature \tex=\trot$\sim$5.2-8~K derived by \citet{muller-etal06} 
and \citet{wiklind+combes96} for
molecular diagnostics of the physical conditions of dense cores such as HCO$^+$. 
Using these same ($n$, \tkin), we 
in turn obtain \tex\ for CO and C of $\sim$6-20 and 8-25~K.
These slightly higher \tex\ values
are consistent with the lower Einstein $A$-coefficients of CO and C compared to HCO$^+$.
\Nco\ and \Nc\ are then 
computed from, e.g., Eq.~2 of \citet{muller-etal06},
and are given in Table~\ref{results tab}.
The largest \Nc\ is 
consistent with the upper limit 
of $\sim10^{18}$~\cs\ reported by \citet{gerin-etal97},
and the CO column density we obtain for the $\sim$7~\kms\ component, 
0.9$^{+3.9}_{-0.5}\times10^{18}$~\cs, is also in
agreement with the estimate 
given by these authors.
Our results yield \NcNco$\sim$2.5 for the two components at
$V_{\rm hel,0}\sim-7$ and $\sim$6~\kms, and an upper limit of $\sim$0.5 for the third component at
$V_{\rm hel,0}\sim-19$~\kms.

Finally, we derived an upper limit on $N_{\rm CO}$ in the NE component of
$\lesssim3.5\times10^{16}$~\cs, assuming a line width of 15~\kms\ 
\citep{muller-etal06}. This is consistent with the smallest $N_{\rm CO}$
of $3\times10^{16}$~\cs\ expected from the range of HCO$^+$ to CO
column density ratios 
determined from our data and that of \citet{muller-etal06}.
Non-detection of the targeted molecules in the NE component could be due to the
change in absorption reported by \citet{muller+guelin08} in this source.


\section{Discussion and conclusion} 
\label{sec:discussion}

Observations of C, C$^+$ and CO can be used
to probe the physical conditions of the gas. 
However, studies of the (clumpy) interstellar medium in external galaxies are 
complicated by several issues: 
(i) the large observing beams which encompass several regions
of differing conditions;
(ii) the difficulty of performing [\ci] observations due to the weakness
of the lines and the poorer atmospheric transparency 
at the high frequencies of the [\ci] transitions (492 and 809~GHz); and
(iii) the impossibility to access the [\cii] 158-$\mu$m transition from the ground.
The observations presented here make it possible to tackle the first two points.
Indeed, since the lines are detected in absorption 
and the galaxy responsible for the latter is almost face-on, 
the line profiles will not be affected much by rotation 
broadening.
Regarding point (ii), it is worthy to note that in fact,
no [\ci] absorption has yet been detected in a local galaxy at $z\sim0$.
For \pks,
the absorbing material is at a redshift of $\sim$0.89, shifting the
492-GHz [\ci] line to $\sim$261~GHz where the atmosphere is more transparent
and stable. 

While absorption data circumvent some of the problems inherent to emission
spectra, some limitations unfortunately remain, in particular,
our (lack of) knowledge of the structure of the absorbing gas.
This translates partly into the estimation of the covering factor $f_{\rm c}$, which is
one of the major sources of uncertainty (along with \tex) in the determination 
of the column densities.
For CO(4$-$3), the largest optical depth is $\sim$2, meaning that the
absorption almost reaches the zero intensity, but it is not heavily
saturated as indicated by the non-detection of the less abundant $^{13}$CO.
Indeed, \citet{muller-etal06} derive $^{12}$C/$^{13}$C 
$\sim$27, so that if CO were heavily saturated,
the expected optical depth of 
$^{13}$CO would be $\gtrsim$0.1, i.e. at the 2-$\sigma$ detection limit.
Hence, the covering factor of the absorbing material must 
be close to unity, as noted by previous studies of this source. 
This is consistent with the picture of individual clouds with sizes
of order $\sim$1~pc (cf. 
\citealt{menten-etal99,muller+guelin08}, and sizes of typical Galactic molecular clouds
of up to a few pc)
in front of a background source of projected size $\lesssim$1~pc
at the distance of the galaxy,
corresponding to the deconvolved core size of 
$\sim0.228\times0.148$~mas measured by \citet{jin-etal03} 
for the SW image of \pks\ at 7~mm with the VLBA.

Comparing the total H column density derived with different methods
can indicate whether the ``cloud'' is uniform
\citep[e.g.][]{wiklind+combes97}.
For the SW source,
there is no strong evidence for clumpiness, since in this case
the total H column density derived from X-ray data 
($\sim3.5\times10^{22}$~\cs, \citealt{mathur+nair97}) 
is comparable to that derived from millimeter CO absorption
($\sim5\times10^{22}$~\cs, assuming CO/H$_2$$\sim5\times10^{-5}$),
or from \ci\ absorption ($\sim6\times10^{22}$~\cs, assuming a typical 
C/H$_2$$\sim8\times10^{-5}$, \citealt{frerking-etal89}). 
Hence, the assumption that the individual velocity components
consist of material with homogeneous physical and chemical
conditions seems sufficient, which allows us to compare the
column density ratios.\\
 
Our data indicate two types of \NcNco\ (see Table 3 for comparisons):
a low value, $\lesssim$0.5, representative of dense cores or PDRs ($\lesssim0.5-1$),
and a somewhat high value, $\sim$2.5, close to those derived
for translucent clouds ($\sim3-6$). 
Qualitatively, low \NcNco\ are found in environments 
with low neutral atomic carbon abundances,
such as high-UV environments (where C becomes ionized), 
or high-density environments (where CO formation is efficient).
 The presence of high density tracers such as HCO$^+$, HCN and CS
supports the latter case of a dense
core for the $V_{\rm hel}=-19$~\kms\ component.
On the other hand, high \NcNco\ are representative of low-density, 
low-column density, mild-UV environments 
where CO is photodissociated but atomic carbon is not ionized.
 For the $V_{\rm hel}\sim-7$~\kms\
component, the detection of hot NH$_3$ ($V_{\rm LSR}\sim$5~\kms, \citealt{henkel-etal08}) and
large \hi\ optical depth \citep[e.g.][]{koopmans+debruyn05} point toward a low density PDR. 
Since hot NH$_3$ was not detected at $V_{\rm hel}\sim5$~\kms, the high \NcNco\
here more likely indicates diffuse or translucent conditions. 
An absorbing source at $z\sim2.4$ with similar diffuse material, 
albeit a higher \NcNco\ of $\sim$10, was reported by \citet{srianand-etal08}.

Atomic C variations could also be explained
by chemical evolution \citep{maezawa-etal99}: 
since the timescale for conversion of C to CO is comparable 
to the dynamical timescale of a dense core ($\sim10^6$~yr),
\NcNco\ is expected to be high in the early stage of dense core
formation and to decrease with time. Hence,
our \NcNco$\sim$2.5 could be indicative of a cloud in an intermediate stage 
of evolution from diffuse to dense gas, where C 
has not completely converted into CO.\\

In conclusion, 
the data presented here demonstrate the ability
of the eSMA to provide observers with high angular resolution,
high sensitivity observations.
Furthermore, they allowed us to investigate the physical and
chemical conditions of the material obscuring the SW component
of \pks, and determine \NcNco\  directly in a dense molecular cloud
located at $z$=0.886.

\acknowledgements
We are very much
indebted to Ray Blundell, Gary Davis and Tom Phillips, 
the directors of the SMA, JCMT and CSO, for their
continued support and without whom the eSMA would not be.
The development of the eSMA has been facilitated by grant 614.061.416
from the Netherlands Organization for Scientific Research, NWO  
and NSF grant AST-0540882 to the Caltech  
Submillimeter Observatory.
AMH is supported by a
National Science Foundation Graduate Research Fellowship.



\clearpage

\begin{table}[ht]
\caption{Main observational parameters for \pks \label{tab:obs}}
\begin{center}
\begin{tabular}{lccc}
\hline
\hline
Parameter                    & DSB continuum              & C ($^3$P$_1-^3$P$_0$) & $^{12}$CO(4-3)   \\
\hline                                                                               
$\nu_{\rm rest}$ (GHz)\tablenotemark{a}   & --                     & 492.16065                 & 461.04077        \\
$\nu_{\rm obs}$ (GHz)\tablenotemark{a}    & 1.1~mm                 & 260.97965                 & 244.47761        \\
Channel width           & 2$\times$$\sim$1.5 GHz\tablenotemark{b}  & 0.93~\kms\tablenotemark{c}              & 0.99~\kms\tablenotemark{c}        \\
Beam FWHM              & $0\farcs55 \times 0\farcs22$ & $0\farcs62 \times 0\farcs30$ & $0\farcs57 \times 0\farcs40$   \\
Beam P.A.                    & $32^\circ$             & $38^\circ$                & $15^\circ$       \\
Array                        & eSMA                   & eSMA                      & SMA              \\
\hline
\end{tabular}
\end{center}
\tablenotetext{a}{The rest and observing frequencies ($\nu_{\rm rest}$ and $\nu_{\rm obs}$) 
are related by $\nu_{\rm obs} = \nu_{\rm rest} / (1+z)$, with $z$ = 0.88582.}
\tablenotetext{b}{Note that in the 230\,GHz band, the eSMA bandwidth is limited 
by that of the JCMT's receiver A3
--- see text for details.}
\tablenotetext{c}{Corresponding to a uniform frequency resolution of $\sim$0.8~MHz.}
\end{table}

\begin{table}[ht]
\caption{Observed and derived line parameters for the SW component \label{results tab} } 
\begin{scriptsize}
\begin{center}
\begin{tabular}{lrcrcrr} 
\hline\hline 
Transition & \multicolumn{1}{c}{$V_{\rm hel,0}$}
 & \multicolumn{1}{c}{$\tau_0$} & \multicolumn{1}{c}{FWHM} 
 & $\int \tau$d$V$ &  \multicolumn{1}{c}{$N$\tablenotemark{a}} 
 & \multicolumn{1}{c}{$N_{\rm C}$} \\ 
\cline{7-7}
	&   \multicolumn{1}{c}{(\kms)}
  &     & \multicolumn{1}{c}{(\kms)} &  (\kms) & \multicolumn{1}{c}{(10$^{17}$  cm$^{-2}$)} 
 & \multicolumn{1}{c}{$N_{\rm CO}$} \\ 
\hline 
[\ci]        &  -19.0 & $<$0.15\tablenotemark{c} & 10.0\tablenotemark{d} & $<$1.6 & $<$1.6 & $<$0.6 \\
	     &   -7.8 &  0.65   & 15.1   & 10.3 & 10.2$^{+4.5}_{-4.4}$\phn & 2.7$^{+0.5}_{-2.0}$\\ 
	     &    5.0 &  0.99   & 12.9   & 13.6 & 13.5$^{+6.0}_{-5.8}$\phn & 2.2$^{+0.4}_{-1.6}$ \\
CO 	     &-19.0\tablenotemark{b} &  0.86   & 13.7   & 12.5 & 2.6$^{+10.9}_{-\phn1.3}$ & \multicolumn{1}{r}{--\phn} \\
	     & -6.5\tablenotemark{b} &  1.59   & 11.0\tablenotemark{b} & 18.6 & 3.8$^{+16.2}_{-\phn1.9}$ & \multicolumn{1}{r}{--\phn} \\
	     &  6.9\tablenotemark{b} &  2.21   & 13.0\tablenotemark{b} & 30.6 & 6.2$^{+27.1}_{-\phn3.2}$ & \multicolumn{1}{r}{--\phn} \\
$^{13}$CO, C$^{18}$O &  \multicolumn{1}{r}{--\phn} & $<$0.1\tablenotemark{c} & 10.0\tablenotemark{d} & $<$1.1 & $<$0.2 
& \multicolumn{1}{r}{--\phn} \\ 
\hline 
\end{tabular} 
\end{center}
\tablenotetext{a}{Using $T_{\rm ex}$=10$^{+10}_{-4}$ K for CO, $^{13}$CO, C$^{18}$O, 
and 15$^{+10}_{-7}$ K for \ci.}
\tablenotetext{b}{These positions/FWHM were constrained to the positions obtained from 
the fit to the \ci\ transition (within the errors) and/or to the positions/FWHM
obtained by \citet{muller-etal06}.}
\tablenotetext{c}{2$\sigma$ upper limit in $\sim$6~\kms\ channel.}
\tablenotetext{d}{Fixed.}
\end{scriptsize}
\end{table}

\begin{table}
\caption{Comparison of \NcNco\ measurements.\label{tab:ratios}}
\begin{scriptsize}
\begin{center}
\begin{tabular}{lcl}
\hline\hline 
Source & \NcNco & References\\
\hline
Translucent clouds& $\sim$3-6 & \citealt{stark+vandishoeck94} \\ 
Diffuse clouds & $\sim$1-80 & \citealt{federman-etal80} \\
Diffuse material at $z$=2.418 & $\sim$10 & \citealt{srianand-etal08} \\
\hline
\multicolumn{3}{c}{Molecular clouds} \\
\hline
Ophiucus & $\sim$0.05-1 & \citealt{frerking-etal89} \\
G34.3+0.2 & $\sim$0.1-1 & \citealt{little-etal94} \\
W3, NCG2024, S140, Cep A & $\sim$0.1-0.5 & \citealt{plume-etal99} \\
TMC1, L1527 & $\sim$0.06-0.25 & \citealt{maezawa-etal99} \\
Orion A/B (multi position) & 0.05-0.21 & \citealt{ikeda-etal02} \\ 
\hline
\multicolumn{3}{c}{PDRs (multi positions)} \\
\hline
M17, S140; IC 348 & 0.02-0.25 & \citealt{keene-etal85,sun-etal08} \\
NGC1977 & 0.03-0.2 & \citealt{minchin+white95} \\
Orion Bar & 0.04-0.3 & \citealt{jansen-etal95} \\
S106 & 0.02-0.09 & \citealt{schneider-etal03} \\
IC 63 (high/low density) & 0.1-0.7 / 2-3 & \citealt{jansen-etal96} \\
\hline
\multicolumn{3}{c}{Galaxies} \\
\hline
M33 (4 positions)  & 0.03-0.16 & \citealt{wilson97} \\ 
IC 342 (center) & 0.5-0.6 & \citealt{israel+baas03} \\ 
NGC 253; Maffei 2 (center) & 0.2-0.3 & \citealt{israel-etal95,israel+baas03} \\ 
M82 & $\sim$0.5 & \citealt{white-etal94} \\ 
M83 (5$''$, 5$''$) & $\sim$0.3 & \citealt{petitpas+wilson98} \\
M83 (center), NGC 6946  & $\sim$0.9 & \citealt{israel+baas01} \\ 
M51 (center)   & 0.4-1.0 & \citealt{israel-etal06} \\ 
Spiral at $z$=0.886 & $\lesssim$0.5, $\sim$2.5 & This work \\
\hline
\end{tabular}
\end{center}
\end{scriptsize}
\end{table}

\clearpage

\vfill

\vfill

\end{document}